\documentclass[letterpaper,11pt]{article}
\usepackage{graphicx}
\usepackage{comment}
\usepackage{url}

\begin{document}

\newcommand{\singlespace}{\renewcommand{\baselinestretch}{1} \normalsize}
\newcommand{\doublespace}{\renewcommand{\baselinestretch}{1.1} \normalsize}

\title{Computational Study of Evolutionary Selection Pressure on
Rainbow Trout Estrogen Receptors}

\author{Conrad Shyu\\
\url{conrads@uidaho.edu}\\
 \\
Department of Physics\\
University of Idaho\\
Moscow, ID 83844-0903\\
\and
Celeste J. Brown\\
\url{celesteb@uidaho.edu}\\
 \\
Department of Biological Sciences\\
University of Idaho\\
Moscow, ID 83844-3051\\
\and
F. Marty Ytreberg\\
\url{ytreberg@uidaho.edu}\\
 \\
Department of Physics\\
University of Idaho\\
Moscow, ID 83844-0903}

\maketitle \doublespace

%
\begin{abstract}
Molecular dynamics simulations were used to determine the binding affinities
between the hormone 17$\beta$-estradiol (E2) and different estrogen receptor
(ER) isoforms in the rainbow trout, \emph{Oncorhynchus mykiss}. Previous
phylogenetic analysis indicates that a whole genome duplication prior to the
divergence of ray-finned fish led to two distinct ER$\beta$ isoforms, ER$\beta
1$ and ER$\beta 2$, and the recent whole genome duplication in the ancestral
salmonid created two ER$\alpha$ isoforms, ER$\alpha 1$ and ER$\alpha 2$. The
objective of our computational studies is to provide insight into the
underlying evolutionary pressures on these isoforms. For the ER$\alpha$ subtype
our results show that E2 binds preferentially to ER$\alpha 1$ over ER$\alpha
2$. Tests of lineage specific $d$N/$d$S ratios indicate that the ligand binding
domain of the ER$\alpha 2$ gene is evolving under relaxed selection relative to
all other ER$\alpha$ genes. Comparison with the highly conserved DNA binding
domain suggests that ER$\alpha 2$ may be undergoing neofunctionalization
possibly by binding to another ligand. By contrast, both ER$\beta 1$ and
ER$\beta 2$ bind similarly to E2 and the best fitting model of selection
indicates that the ligand binding domain of all ER$\beta$ genes are evolving
under the same level of purifying selection, comparable to ER$\alpha 1$.
\end{abstract}

%
\section{Introduction}
Estrogens are essential endogenous hormones that modulate the development and
homeostasis of a wide range of target tissues, such as the reproductive tracts,
breast and skeletal system \cite{ruff00}. Estrogenic hormones have
multi-faceted and wide-ranging effects in vertebrate animals. For estrogens
such as 17$\beta$-estradiol (E2) to exert their biological effects, they must
interact with cellular estrogen receptors (ER). Studies have shown that ERs are
part of two distinct estrogenic transduction pathways. One pathway provides a
rapid, nongenomic pathway initiated by membrane bound ERs at the cell surface
\cite{ruff00, nagler00, pakdel00}. The other pathway provides direct genomic
control in which ERs act as transcription factors within the cell nucleus
\cite{froehlicher09, menuet04}. These ERs are members of the nuclear receptor
superfamily of ligand-modulated transcription factors \cite{drean95, nagler07,
tsai94}. There are two different subtypes of these ERs, referred to as $\alpha$
and $\beta$, each encoded by a separate gene.

\begin{figure}[tb]
\begin{center}
\includegraphics[width=3.6in]{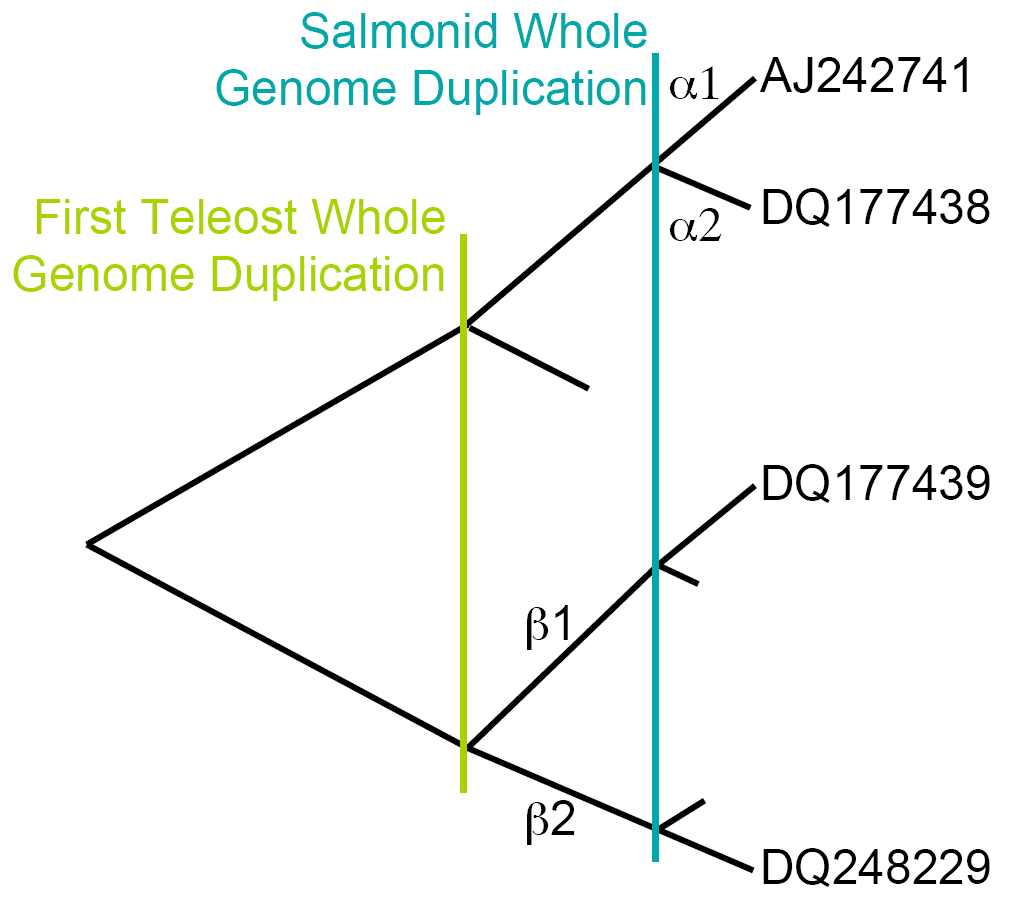}
\end{center}
\caption[Schematic showing the inferred evolutionary history of the
\emph{Oncorhynchus mykiss} estrogen receptors]{Schematic showing the inferred
evolutionary history of the \emph{Oncorhynchus mykiss} estrogen receptors.
Vertical bars mark inferred whole genome duplication events; short branches
mark inferred duplications that were lost over time.} \label{fig:tree}
\end{figure}

%
%
Recently, Nagler et al \cite{nagler07} reported the novel ER$\alpha 2$ and both
ER$\beta$ isoforms in the rainbow trout, \emph{Oncorhynchus mykiss}, and
performed a comprehensive phylogenetic analysis with all other known fish ER
gene sequences. Their phylogenetic analysis indicates that the duplication
leading to the two ER$\beta$ isoforms arose prior to the divergence of the ray
finned fish attributable to a whole genome duplication that occurred in the
Teleost ancestor (see Fig \ref{fig:tree}) \cite{hoegg04}. The ER$\alpha$
isoforms, on the other hand, appear to have arisen as a result of a second more
recent whole genome duplication event that occurred in the salmonid ancestor
25--100 million years ago \cite{allendorf84}. These results indicate that the
second ER$\alpha$ isozyme that arose during the earlier genome duplication
appears to have been lost subsequently, since no other ray finned fish are
known to have a second ER$\alpha$ isoform. This also indicates that the
expected duplications of ER$\beta 1$ and ER$\beta 2$ were lost subsequent to
the salmonid genome duplication.

%
%
The purpose of the study is to employ molecular dynamics simulations to
determine the binding affinities between E2 and ERs of the different isoforms
in the rainbow trout and to use the results to provide insight into the
underlying evolutionary selection pressure on the ERs. Our binding affinity
results obtained from insertion and deletion are very similar indicating that
our simulations are well converged and that accurate estimates of binding
affinities were obtained. Our results show that E2 binds preferentially to
ER$\alpha 1$ over ER$\alpha 2$. By contrast, the difference in binding affinity
is less significant for the $\beta$ subtype, i.e., both isoforms bind similarly
to E2. We also computed $d$N/$d$S ratios for the ER isoforms. These results
suggest that the ER$\alpha 1$ gene is evolving under relaxed selection compared
to all other salmonid ER$\alpha$ genes.

\section*{Materials and Methods}
\subsection*{Receptor Structures}
\begin{figure}[tb]
\begin{center}
\includegraphics[width=3.6in]{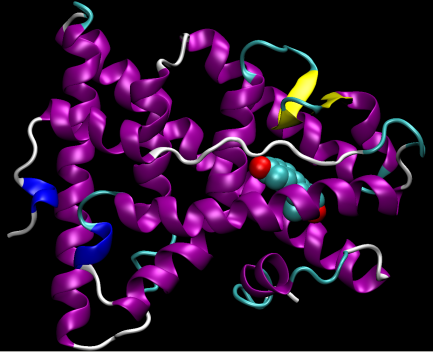}
\end{center}
\caption[Crystal structure of human estrogen receptor binding domain bound to
the hormone 17$\beta$-estradiol]{Crystal structure of human estrogen receptor
binding domain bound to the hormone 17$\beta$-estradiol. Similar human ER
structures were used as templates to generate structures for trout ERs. Image
was rendered using VMD \cite{humphrey96}.} \label{fig:1qku}
\end{figure}

The initial coordinates for the estradiol were first extracted from the human
ER-E2 complexes (PDB: 1QKU) (Fig \ref{fig:1qku}). The topologies were then
generated by the PRODRG server \cite{schuttelkopf04} with the options of full
charges and no energy minimization. The rainbow trout ER \emph{holo} structures
for the E2 binding domain were generated by SWISS-MODEL \cite{arnold06} using
human ER as templates (PDB entries 1A52 for ER$\alpha$'s and 3ERT for
ER$\beta$'s). Sequence identities between trout and human estrogen binding
domains are within the range of 75--85\%. The estradiol was first docked into
the binding pocket of the receptor \emph{holo} structure with AutoDock
\cite{morris98}. In this protocol, the receptor structure is held rigid and the
estradiol is free to rotate and explore most probable binding poses using the
Lamarckian genetic algorithm. The number of genetic algorithm runs was set to
1,000 with a population size of 5,000 individuals and 5,000,000 generations.
The number of evaluations was set to 2,500,000 for each individual in the
population to ensure thorough exploration of the search space. The mutation
rate was set to 0.02 and crossover 0.8. Two-point crossover was used to
generate the offspring at each successive generation. The genetic algorithm
automatically preserved the 10 best-fit individuals to the next generation and
the 10 least-fit individuals were not used to generate offspring. A total of
1,000 independent docking trials were performed for each of the four ERs. The
best binding pose from each trial was collected and ranked based on the scores.
These best-fit binding poses were first visually inspected for consistency with
human ER and the one with the highest score was then used as the starting
structure for the simulations.

\subsection*{Thermodynamic Cycle}
To estimate E2-ER binding affinities we note that, since the free energy is a
state function, it permits the selection of an arbitrary path connecting the
bound and unbound states. Therefore, we decomposed the binding free energy
calculation into several steps in which the E2 is annihilated (i.e., decoupled)
from its bound state in the receptor complex and then made to reappear in
solution to complete the thermodynamic cycle. For brevity, we subsequently
define \emph{deletion} to be when interactions between E2 and its environment
are turned off and \emph{insertion} to be when these interactions are turned
on.

\begin{figure}[tb]
\begin{center}
\includegraphics[width=3.8in]{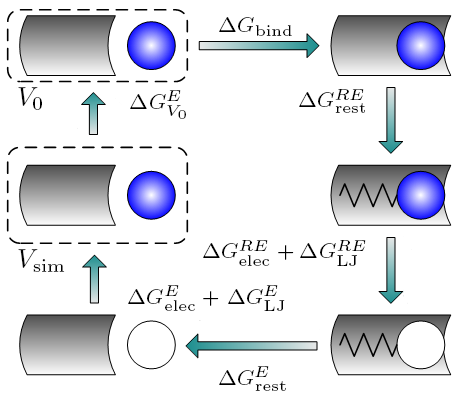}
\end{center}
\caption[Thermodynamic cycle used for calculating binding affinities]
{Thermodynamic cycle used for calculating binding affinities. Since the free
energy is a state function, the calculation of binding affinity is decomposed
into several steps \cite{chipot07}. Eqn \ref{equ:total} was used to calculate
the binding affinity $\Delta G _{\rm bind}$ between the ERs and E2. The gray
curved rectangle represents the receptor, the blue circle represents E2 with
all interactions turned on, and the white circle indicates that all
interactions are turned off. The spring represents the restraints between E2
and receptors.} \label{fig:cycle}
\end{figure}

Fig \ref{fig:cycle} shows the thermodynamic cycle we used to calculate binding
affinities (see also Refs \cite{kirkwood35, mitchell91, mobley06, shirts03,
shirts05}). Starting with upper right schematic and moving clockwise, the fully
interacting E2 (blue) is first restrained in the binding pocket of the
receptor. Here \emph{RE} represents the solvated complex of the E2 and
receptor, and $\Delta G ^{RE} _{\rm rest}$ denotes the free energy of
restraining the E2 in the binding pocket of the receptor which will depend on
the details of restraint. Next, the electrostatic and Lennard-Jones
interactions of the E2 are gradually turned off (white) in two separate steps
using alchemical simulations. The free energies $\Delta G ^{RE} _{\rm elec}$
and $\Delta G ^{RE} _{\rm LJ}$ are associated with deleting or inserting
electrostatic and Lennard-Jones interactions respectively. With the E2 fully
decoupled from its environment, the restraint is then removed. The free energy
$\Delta G ^{E} _{\rm rest}$ is associated with the removal of this restraint.
Next the E2 interactions are turned back on with no receptor present. The free
energy $\Delta G ^{E} _{\rm elec}$ and $\Delta G ^{E} _{\rm LJ}$ are associated
with turning on the electrostatic and Lennard-Jones interactions respectively.
Finally, we account for the difference between the standard ($V _0$) and
simulation volume ($V _{\rm sim}$). The binding affinity between the estrogen
receptors and E2 is thus the sum of the free energies, \begin{equation} -
\Delta G _{\rm bind} = \Delta G ^{RE} _{\rm rest} + \Delta G ^{RE} _{\rm elec}
+ \Delta G ^{RE} _{\rm LJ} + \Delta G ^{E} _{\rm rest} + \Delta G ^{E} _{\rm
elec} + \Delta G ^{E} _{\rm LJ} + \Delta G ^{E} _{V _0}. \label{equ:total}
\end{equation}

\begin{figure}[tb]
\begin{center}
\includegraphics[width=4.0in]{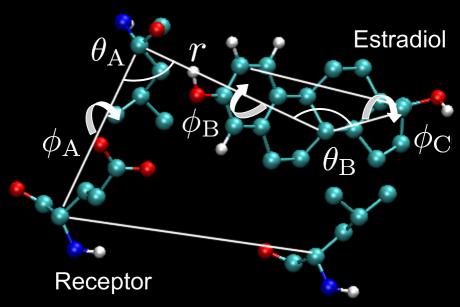}
\end{center}
\caption[Restraints used for 17$\beta$-estradiol]{Restraints used for
17$\beta$-estradiol. We used six harmonic restraints acting on three anchor
atoms in the E2 and three in the receptor. The restraints consisted of one
distance ($r$), two angle ($\theta _{\rm A}$ and $\theta _{\rm B}$), and three
dihedral ($\phi _{\rm A}$, $\phi _{\rm B}$, and $\phi _{\rm C}$) restraints
that determine the orientation of three carbons in the E2 relative to three
$\alpha$-carbons in the receptors.} \label{fig:restraint}
\end{figure}

%
\subsection*{Restraints}
To facilitate convergence restraints were applied to restrict the positions of
E2 relative to the receptors. Boresch et al \cite{boresch03} and Mobley et al
\cite{mobley06} reported that the presence of multiple metastable ligand
orientations can cause convergence problems for free energy estimates. The
authors further suggested using a restraining potential to keep the ligand in
the binding site during the simulation process. With such a restraining
potential the ligand is no longer required to sample the entire simulation
volume (particularly a problem when ligand is decoupled). Moreover, the
restraint minimizes the detrimental effects of end-point singularities commonly
reported in alchemical simulations \cite{mitchell91, mobley06, shirts05}.
Mobley et al \cite{mobley06} also pointed out that the equilibrium geometry of
the restraints is arbitrary and will not affect the asymptotic estimate of the
binding free energy. In this work, we judiciously selected anchor atoms from
the more rigid alpha helices that form the E2 binding pocket. The restraints
included one distance (with the force constant of 1000 kJ/mol/$\rm{nm}^2$), two
angle (1000 kJ/mol/rad), and three dihedral restraints (1000
kJ/mol/$\rm{rad}^2$) that determine the orientation of three carbons in the E2
relative to three $\alpha$-carbons in the receptors (see Fig
\ref{fig:restraint}).

\subsection*{Simulation Protocols}
All simulations were performed with the GROMACS 4.0 \cite{hess08} compiled in
single-precision mode at a constant temperature of 277 K in a periodic box with
an edge length of approximately 8.2 nm and the default GROMOS-96 43A1
forcefield \cite{gunsteren96}. The simulation systems each contained
approximately 16,500 Simple Point Charge (SPC) water molecules
\cite{berendsen81}. Short-range interactions were evaluated using a neighbor
list of 1.0 nm updated at every 10 steps. Van der Waals interactions used a
cutoff with a smoothing function such that the interactions slowly decayed to
zero between 0.75 nm and 0.90 nm. A long-range analytical dispersion correction
was applied to the energy and pressure to account for the truncation of the
Lennard-Jones interactions \cite{allen89}. Electrostatic interactions were
evaluated using the particle mesh Ewald (PME) \cite{darden99} with a real space
cutoff of 1.0 nm, a spline order of 6, a Fourier spacing of 0.1 m, and relative
tolerance between long and short range energies of $10 ^{-6}$. All bonds to
hydrogen were constrained with LINCS \cite{hess97} with an order of 12, and a
time step of 2 fs was used for dynamics.

For equilibration, the systems were first minimized using 1,000 steps of L-BFGS
(Broyden-Fletcher-Goldfarb-Shanno) \cite{broyden70}, followed by 1,000 steps of
steepest descent minimization. The system was then subject to 1.0 ns of
simulation using isothermal molecular dynamics. This was followed by another
1.0 ns of simulation using isothermal-isobaric dynamics with the Berendsen
barostat with a time constant of 1.0 ns. For all simulations the temperature
was maintained at 277 K using Langevin dynamics \cite{berendsen88} with a
friction coefficient of 1.0 amu/ps. The coupling time was set to 0.5 ps, and
the isothermal compressibility was set to $4.5 \times 10 ^{-5}$ bar$^{-1}$.

After equilibration, production simulations were run with isothermal-isobaric
conditions using Langevin dynamics at the temperature of 277 K. The pressure
was maintained at 1.0 atm using the Parrinello-Rahman algorithm \cite{laio02}.
The temperature was chosen as it closely resembles the water temperature for
the natural habitat of rainbow trout. Energies were recorded every 0.2 ps
during production runs, and trajectory snapshots every 1.0 ps. The first 50\%
of each simulation was discarded for equilibration.

\subsection*{Free Energy Calculations}
We used the formula suggested by Boresch et al \cite{boresch03} to analytically
calculate the free energy $\Delta G ^{E} _{\rm rest}$ associated with adding
the restraints to E2 when decoupled from its environment. We also analytically
calculated the free energy $\Delta G ^{E} _{V _0}$ that accounts for the
difference between the standard ($V _0$) and simulation volume ($V _{\rm sim}$)
\cite{boresch03}.

The free energies $\Delta G ^{RE} _{\rm elec}$, $\Delta G ^{RE} _{\rm LJ}$,
$\Delta G ^{E} _{\rm elec}$, and $\Delta G ^{E} _{\rm LJ}$, were estimated
using the thermodynamic integration (TI) method \cite{kirkwood35, shirts05,
shirts03}. To minimize the numerical integration errors we employed the
polynomial regression techniques to calculate free energy difference, instead
of trapezoidal quadrature \cite{shyu09}. Separate simulations were performed
for changes in the Lennard-Jones with 21 values of the scaling parameter,
$\lambda$ = 0.0, 0.05, 0.1 \ldots 0.9, 0.95, and 1.0, and the electrostatics
with 11 $\lambda$ values, $\lambda$ = 0.0, 0.024, 0.095, 0.206, 0.345, 0.5,
0.655, 0.794, 0.905, 0.976, and 1.0. For simulations with only Lennard-Jones,
all partial charges were set to zero and the soft-core scaling parameter was
set to 0.5. Once the neutral atoms were fully grown in the solvent, the second
simulations then computed the free energy associated with the electrostatics
with a soft-core scaling parameter of 0.0. This was accomplished by increasing
the partial charges from zero to their final values given by the forcefield.

The free energy associated with the restraints, $\Delta G ^{RE} _{\rm rest}$
was calculated using the Bennett acceptance ratio approach \cite{bennett76}. We
performed 1.0 ns equilibrium simulation for the estradiol-receptor complex
using each of the harmonic restraining potentials with force constants of 0,
25, 40, 60, 90, 150, 200, 300, 450, 700, and 1000 kJ/mol/$\rm{nm}^2$ for
distance, kJ/mol/rad for angle, and kJ/mol/$\rm{rad}^2$ for dihedral
restraints. The first 0.5 ns of each simulation was discarded for equilibration
and the remaining 0.5 ns was used to compute the free energy differences. No
attempt was made to optimize the efficiency of the calculation since our
primary objective was to obtain accurate estimates of the restraining free
energies.

\subsection*{Evolutionary Analyses}
The following sequences were extracted from GenBank: AB037185, AF349412,
A133920050, AY727528, AY775183, BD105560, AB190289, AJ487687, AY055725,
AF061275, AF253505, AY520443, AJ242741, DQ009007, DQ248228, DQ177438, X89959,
TNU7560, AY422089, AF298183, AF136979, AY074780, AB007453, AJ006039, AF253062,
AY223902, ORZMER, AY917147, AF326201, AY305026, NM\_180966, NM\_174862,
AB003356, AB070630, AB070901, AB083064, AB117930, AB190290, AF061269, AF136980,
AF177465, AF185568, AF298181, AF298182, AF349413, AF349414, AF516874, AJ275911,
AJ289883, AJ314602, AJ314603, AJ414566, AJ414567, AJ489523, AJ580050, AY074779,
AY211021, AY211022, AY305027, AY307098, AY508959, AY566178, AY770578, AY917148,
BC044349, BC086848, DQ177439, DQ248229, TNU75605. The first 30 are ER$\alpha$
sequences and the other 39 are ER$\beta$ sequences, and the following analysis
was done separately for these two subtypes. The codons were aligned based upon
their aligned amino acid sequences, and these alignments were used to infer
tree topologies using the neighbor joining method. Then the ligand binding
domains were extracted from the alignments. PAML was used to test several
codon-based likelihood models that allow for variable $d$N/$d$S ratios among
lineages based upon the inferred phylogenies and the aligned ligand binding
domains \cite{yang07}.

%
%
\section*{Results and Discussion}
Experimental binding affinity results are not readily available for the four
trout ERs due to the difficulty in isolating the different isoforms. Thus, to
verify our methodology for estimating $\Delta G _{\rm bind}$ for rainbow trout
ER-E2 we first performed simulations using human ER (PDB: 1QKU) at 300 K and
compared the binding affinities to the experimental results. Our computational
estimates at 300 K are $-63.2$ kJ/mol for insertion (when interactions between
E2 and its environment are turned on) and $-59.6$ kJ/mol for deletion (when
interactions between E2 and its environment are turned off). Experimental
binding affinity for human ER is $-52.3$ kJ/mol at 300 K \cite{petit95}. Thus,
our human ER binding affinity estimates are within about 10 kJ/mol of
experiment which is within the expected error due to the atomic models
\cite{hess08}. The trout ER simulations followed exactly the same procedure as
human, beginning with docking the E2 into the ER. It is important to note that
our ER simulations were performed at 277 K to closely mimic the water
temperature of rainbow trout natural habitat.

Table \ref{tab:summary} shows our binding affinity results from both insertion
and deletion. Simulation results from both deletion and insertion of
electrostatics and Lennard-Jones interactions provide a rudimentary assessment
of the accuracy of our calculations (note that there may be inaccuracies in the
atomic models but that is beyond the scope of this study). The fact that both
insertion and deletion give very similar results strongly suggests that our
simulations are well converged and that accurate estimates of binding
affinities have been obtained.

\begin{table}[tb]
\begin{center}
\caption[Estrogen receptor binding affinities for different isoforms obtained
at 277 K]{Estrogen receptor binding affinities for different isoforms obtained
at 277 K. All results are in kJ/mol. The binding affinities $\Delta G
_{\rm{bind}}$ were calculated using Eqn \ref{equ:total}. Both insertion and
deletion directions give very similar results which demonstrates that our
simulations are well converged.} \label{tab:summary}
\begin{tabular}{r r r r r}
\multicolumn{5}{r}{(A) Insertion} \\
\hline
   & ER$\alpha 1$ & ER$\alpha 2$ & ER$\beta 1$ & ER$\beta 2$ \\
\hline
$\Delta G ^{RE} _{\rm rest}$ & -21.5 & -27.6 & -31.4 & -24.2 \\
$\Delta G ^{RE} _{\rm elec} + \Delta G ^{RE} _{\rm LJ}$ &
-146.7 & -126.5 & -132.4 & -144.1 \\
$\Delta G ^{E} _{\rm rest}$ & 48.9 & 49.9 & 52.4 & 49.9 \\
$\Delta G ^{E} _{\rm elec} + \Delta G ^{E} _{\rm LJ}$ & 61.8 &
\multicolumn{1}{c}{---} & \multicolumn{1}{c}{---} & \multicolumn{1}{c}{---} \\
$\Delta G ^{E} _{V _0}$ & -6.8 &
\multicolumn{1}{c}{---} & \multicolumn{1}{c}{---} & \multicolumn{1}{c}{---} \\
\hline
$\Delta G ^{RE} _{\rm bind}$ &
\textbf{-64.3} & \textbf{-49.2} & \textbf{-56.4} & \textbf{-63.4} \\
\hline
\multicolumn{5}{r}{ } \\
\multicolumn{5}{r}{(B) Deletion} \\
\hline
   & ER$\alpha 1$ & ER$\alpha 2$ & ER$\beta 1$ & ER$\beta 2$ \\
\hline
$\Delta G ^{RE} _{\rm rest}$ & -21.5 & -27.6 & -31.4 & -24.2 \\
$\Delta G ^{RE} _{\rm elec} + \Delta G ^{RE} _{\rm LJ}$ &
-149.2 & -125.2 & -131.1 & -144.7 \\
$\Delta G ^{E} _{\rm rest}$ & 48.9 & 49.9 & 52.4 & 49.9 \\
$\Delta G ^{E} _{\rm elec} + \Delta G ^{E} _{\rm LJ}$ & 62.0 &
\multicolumn{1}{c}{---} & \multicolumn{1}{c}{---} & \multicolumn{1}{c}{---} \\
$\Delta G ^{E} _{V _0}$ & -6.8 &
\multicolumn{1}{c}{---} & \multicolumn{1}{c}{---} & \multicolumn{1}{c}{---} \\
\hline
$\Delta G ^{RE} _{\rm bind}$ &
\textbf{-66.6} & \textbf{-47.7} & \textbf{-54.9} & \textbf{-63.8} \\
\hline
\end{tabular}
\end{center}
\end{table}

Our results in Table \ref{tab:summary} show that the E2 binds preferentially to
the ER$\alpha 1$ isoform of the $\alpha$ subtype that has been found in all
salmonids. The other isoform ER$\alpha 2$, which appears to have arisen during
the recent salmonid whole genome duplication, shares 75.4\% sequence identity
with the ER$\alpha 1$ and thus a large number of substitutions have accumulated
since the initial duplication event. To infer the evolutionary pressures that
led to this amount of divergence in both protein sequence and function, we
examined the lineage specific differences in $d$N/$d$S ratios among the
ER$\alpha$ sequences. We used an alignment of the codons in the ligand binding
domain for all ER$\alpha$ sequences and a phylogeny inferred from the
nucleotide sequence by the neighbor joining method (which did not differ
significantly from the tree in \cite{nagler07}). PAML was used to calculate the
log likelihood values and $d$N/$d$S ratios for each of five hypotheses: a
single ratio for all branches, one ratio for all branches except the branch to
the rainbow trout ER$\alpha 2$, separate ratios for the two ER$\alpha$'s from
rainbow trout and the rest of the tree, separate ratios for the rainbow trout
ER$\alpha 2$, all ER$\alpha 1$ from salmonids and the rest of the tree and the
full model where every branch has its own ratio (see Table \ref{tab:paml}).
Using the Aikaike Information Criterion, the model with two ratios, one for the
branch to the rainbow trout ER$\alpha 2$ and one for all other branches is the
best fitting model. For this model, the $d$N/$d$S ratio for all other branches
was 0.09 whereas the ratio for the ER$\alpha 2$ branch was 0.30. In all tests,
the $d$N/$d$S ratio for the ER$\alpha 2$ branch was about three times greater
than the other salmonid branches. Therefore, the ER$\alpha 2$ ligand binding
domain appears to be evolving under relaxed selection relative to the other
salmonid ER$\alpha 1$ ligand binding domains, which is consistent with the
decreased affinity of this domain for E2. It is also possible that ER$\alpha 2$
was evolving in a neutral fashion for a short time, but then developed a new
function and is now undergoing stronger purifying selection. This possibility
could be explored further if more ER$\alpha$ salmonid gene sequences were made
available.

Our results show that both ER$\beta$ isoforms bind similarly to E2, i.e., the
difference between them in binding affinity is small compared to the difference
between the ER$\alpha$ isoforms (see Table \ref{tab:summary}). The two isoforms
share only 57.6\% sequence identity, having arisen prior to the Teleost
radiation, and the difference in their binding affinity might be expected to be
greater, given this large degree of divergence. We performed a similar analysis
of the $d$N/$d$S ratio for these genes by testing the following models: one
$d$N/$d$S ratio for the whole tree, a $d$N/$d$S ratio for each of isoform
ER$\beta 1$ and ER$\beta 2$, $d$N/$d$S ratios for each of the two rainbow trout
isoforms and for each isoform for all other fish and the full model where every
branch has a different $d$N/$d$S ratio (Table \ref{tab:paml}). The best fitting
model for this comparison was the single $d$N/$d$S ratio (0.07) for the entire
tree, indicating that both ER$\beta$ isoforms are under the same level of
purifying selection. This is also consistent with our results showing that
these two ligand binding domains have similar affinity for E2.

\begin{table}[tb]
\begin{center}
\caption[Results of fitting evolutionary models for differences in $d$N/$d$S
ratios.]{Results of fitting evolutionary models for differences in $d$N/$d$S
ratios. LBD and DBD indicate ligand and DNA binding domains, respectively.
O$\alpha 1$, O$\alpha 2$, O$\beta 1$ and O$\beta 2$ are the \emph{O. mykiss}
ER$\alpha 1$, ER$\alpha 2$, ER$\beta 1$ and ER$\beta 2$ genes, respectively.
S$\alpha 1$ indicates all of the salmonid ER$\alpha 1$ genes. $\beta 1$ and
$\beta 2$ indicate ER$\beta 1$ and ER$\beta 2$ from all fish, respectively.
$np$ is the number of parameters in the model, $\ln L$ is the log likelihood
calculated by PAML, and AIC is the Akaike Information Criterion value
\cite{akaike74}. Models labeled with an asterisk are the best fitting models
based upon the AIC values.} \label{tab:paml}
\begin{tabular}{l r r r c}
ER$\alpha$-LBD & \multicolumn{1}{c}{$np$} & \multicolumn{1}{c}{$\ln L$} &
\multicolumn{1}{c}{AIC} & \\
\hline
$H_0$: Everyone is equal & 59 & -5645.4 & 11409  &\\
$H_1$: O$\alpha 2$ $\neq$ others & 60 & -5641.7 & 11403 & $\ast$ \\
$H_2$: O$\alpha 2$ $\neq$ O$\alpha 1$ $\neq$ others & 61 & -5641.6 & 11405 & \\
$H_3$: O$\alpha 2$ $\neq$ S$\alpha 1$ $\neq$ others & 61 & -5641.5 & 11405 & \\
$H_{\rm Full}$: Everyone is different & 115 & -5610.3 & 11451 & \\
\hline
\multicolumn{5}{c}{ } \\
ER$\alpha$-DBD & \multicolumn{1}{c}{$np$} & \multicolumn{1}{c}{$\ln L$} &
\multicolumn{1}{c}{AIC} & \\
\hline
$H_0$: Everyone is equal & 59 & -1121.0 & 2360 & $\ast$ \\
$H_1$: O$\alpha 2$ $\neq$ others & 60 & -1120.8 & 2362 & \\
$H_2$: O$\alpha 2$ $\neq$ O$\alpha 1$ $\neq$ others & 61 & -1120.7 & 2363 & \\
$H_3$: O$\alpha 2$ $\neq$ S$\alpha 1$ $\neq$ others & 61 & -1120.3 & 2363 & \\
$H_{\rm Full}$: Everyone is different & 115 & -1101.8 & 2434 & \\
\hline
\multicolumn{5}{c}{ } \\
ER$\beta$-LBD & \multicolumn{1}{c}{$np$} & \multicolumn{1}{c}{$\ln L$} &
\multicolumn{1}{c}{AIC} & \\
\hline
$H_0$: Everyone is equal & 77 & -7064.6 & 14283 & $\ast$ \\
$H_1$: $\beta 2$ $\neq$ $\beta 1$ & 78 & -7063.9 & 14284 & \\
$H_2$: O$\beta 2$ $\neq$ O$\beta 1$ $\neq$ $\beta 1$ $\neq$ $\beta 2$ & 80 &
 -7062.1 & 14284 & \\
$H_{\rm Full}$: Everyone is different & 151 & -7005.9 & 14314 & \\
\hline
\end{tabular}
\end{center}
\end{table}

These nuclear ERs have a significant and ubiquitous distribution in the rainbow
trout \cite{nagler07, nagler00}. The levels of transcription differ among the
four genes with one isoform having higher transcript levels in most tissues
than the other isoform. For the ER$\alpha$ isoforms, ER$\alpha 1$ has the
higher transcript levels, and for the ER$\beta$ isoforms, ER$\beta 2$ has the
highest transcript levels \cite{nagler07}. While the correlation between
reduced transcription levels and binding affinity is clear in the ER$\alpha$
isoforms, there seems to be no such correlation for the ER$\beta$ isoforms.
These two isoforms share similar binding affinity, and yet, ER$\beta 1$ has
much lower expression levels than ER$\beta 2$ in juvenile rainbow trout. It is
possible that both ER$\alpha 2$ and ER$\beta 1$ have higher expression levels
at other life stages \cite{nagler07}. Given the age of ER$\beta 1$ and the
equivalent levels of both E2 binding affinity and purifying selection compared
with ER$\beta 2$, this ER clearly continues to have an important role as an
estrogen receptor.

It is not as clear what ER$\alpha 2$'s role is as an estrogen receptor. It's
reduced affinity for E2, low transcript levels and evidence for relaxed
selection suggests that this estrogen receptor may be undergoing
subfunctionalization or neofunctionalization. One indication that ER$\alpha 2$
may be undergoing neofunctionalization is that the DNA binding domain of
ER$\alpha 2$ does not have the degree of sequence variation that the ligand
binding domain has. If the ER$\alpha 2$ was undergoing relaxed selection along
it's entire length, the DNA binding domain would also show indications of
greater amino acid divergence (Table \ref{tab:paml}). It appears that ER$\alpha
2$ is not losing its ability to bind to the canonical estrogen receptor element
even though it is losing affinity for E2. This suggests that this gene may be
undergoing neofunctionalization by binding to some other ligand than E2.

\section*{Conclusions}
Using molecular dynamics simulations we estimated the binding affinities
between the hormone 17$\beta$-estradiol (E2) and different estrogen receptor
(ER) isoforms in the rainbow trout, \emph{Oncorhynchus mykiss}. Our results
show that E2 binds preferentially to ER$\alpha 1$ over ER$\alpha 2$. A recent
genome wide duplication event led to two functional ER$\alpha$ isozymes in
\emph{O. mykiss}. Our evolutionary and functional analyses along with Nagler's
evaluation of transcription levels \cite{nagler07} suggest that the ligand
binding domain of ER$\alpha 2$ has been or is currently evolving under relaxed
selection relative to ER$\alpha 1$. Low sequence divergence of its highly
conserved DNA binding domain suggests that ER$\alpha 2$ is likely undergoing
neofunctionalization, in which it continues to recognize the same estrogen
receptor element in the DNA but may be binding to a different ligand. For the
ER$\beta$ subtype both isoforms bind similarly to E2, in keeping with our
evolutionary analyses that both isoforms of this subtype are evolving under the
same degree of purifying selection.

%
\section*{Acknowledgements}
This research was supported by Idaho NSF-EPSCoR, Bionanoscience (BANTech), and
the Initiative for Bioinformatics and Evolutionary Studies (IBEST) at the
University of Idaho. F.M.Y. and C.J.B. were supported by NIH/NCRR P20RR16448
and P20RR016454. The authors would like to thank James J. Nagler and Tim
Cavileer of University of Idaho for suggesting this study.

%
\singlespace
\bibliographystyle{plain}

\end{document}